\documentclass[reprint,aps,prb,floatfix,showpacs]{revtex4-1}

\usepackage{amsfonts}
\usepackage{amsmath}
\usepackage{amssymb}
\usepackage{graphicx}
\usepackage{times}
\usepackage{hyperref}
\usepackage{epsfig}
\usepackage{epstopdf}
\usepackage{dcolumn}

\newcommand{\bear}{\begin{eqnarray}}
\newcommand{\eear}{\end{eqnarray}}
\newcommand{\be}{\begin{equation}}
\newcommand{\ee}{\end{equation}}
\newcommand{\nn}{\nonumber}
\newcommand{\Eqref}[1]{Eq.~(\ref{#1})}

\newcommand{\ud}{\mathrm{d}}
\newcommand{\im}{\mathrm{i}}
\newcommand{\jm}{\mathrm{j}}
\newcommand{\kb}{k_\mathrm{_B}}

\begin{document}

\title{Determination of Boltzmann constant by equipartition theorem for capacitors}

\author{T.~M.~Mishonov}
\email[E-mail: ]{mishonov@gmail.com}
\affiliation{Laboratory for Measurements of Fundamental Constants,
Faculty of Physics, \\
St.~Clement of Ohrid University at Sofia,\\
5 James Bourchier Blvd., BG-1164 Sofia, Bulgaria}

\author{V.~N.~Gourev}
\affiliation{Department of Atomic Physics, Faculty of Physics,\\St.~Clement of Ohrid University at Sofia,\\5 James Bourchier Blvd., BG-1164 Sofia, Bulgaria}

\author{I.~M.~Dimitrova}
\affiliation{Faculty of Chemical Technologies, University of Chemical Technology and Metallurgy, \\
8 Kliment Ohridski Blvd., BG-1756 Sofia, Bulgaria}

\author{N.~S.~Serafimov}
\affiliation{Department of Telecommunications, Technical University Sofia,\\8 Kliment Ohridski Blvd., BG-1000 Sofia, Bulgaria}

\author{A.~A.~Stefanov}
\affiliation{Faculty of Mathematics, St.~Clement of Ohrid University at Sofia,\\
5 James Bourchier Blvd., BG-1164 Sofia, Bulgaria}
\affiliation{Institute of Mathematics and Informatics, Bulgarian Academy of Sciences, \\
Acad.~Georgi Bonchev Str., Block 8, 1113 Sofia, Bulgaria}

\author{E.~G.~Petkov, A.~M.~Varonov}
\email[E-mail: ]{avaronov@phys.uni-sofia.bg}
\affiliation{Faculty of Physics, St.~Clement of Ohrid University at Sofia,\\
5 James Bourchier Blvd., BG-1164 Sofia, Bulgaria}

\date{25 Dec 2018}

\begin{abstract}
A new experimental set-up for Boltzmann constant measurement is described.
Statistically averaged square of voltage $\left<U^2\right>$ is measured for different
capacitances $C$.
Boltzmann constant is determined by the equipartition theorem
$C\left<U^2\right>=k_{_\mathrm{B}}T$.
For fixed capacitance, voltages could be measured for different temperatures.
The set-up consists of low-noise high frequency operational amplifiers ADA4898-2.
An instrumental amplifier is followed by an inverting amplifier,
square of the voltage is created by an analog multiplier AD633 and finally
the averaged signal is measured by a multimeter.
More than 10 high-school students were able to measure the Boltzmann constant with the experimental set-up in the 5$^\mathrm{th}$ Experimental Physics Olympiad with excellent accuracy compared to the price, conditions and available time for the experiment.
A new derivation of the important for the statistical physics theorems by Nyquist and Callen-Welton is given in an appendix in the level of introductory courses in physics listened by teachers. 
For the understanding of the work of the experimental set-up, it is necessary to know
only the equipartition theorem.
\end{abstract}

\maketitle

\section{Introduction}
\label{sec:intro}

\begin{figure}[h]
\includegraphics[scale=0.3]{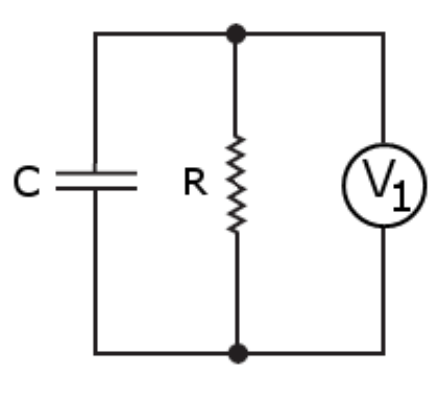}
\caption{Parallely connected capacitor $C$, resistor $R$ and a voltmeter V$_1$ measuring the time dependent thermal fluctuations of the voltage $U(t)$.
A DC voltmeter should show $\left < U \right > =0$ because electric noise does not introduce DC-voltages.}
\label{fig:cr}
\end{figure}

The equipartition theorem~\cite{Herapath,Waterston} describing the relation between the mean thermal energy of a quadratic degree of freedom
\be
\frac12 \left < m v_x^2 \right > = \frac12 \left < kx^2 \right > = \frac12 \left < \tilde I \omega^2 \right >=
\frac12 \left < CU^2 \right > =\! \dots \!= \frac12 \kb T
\nn
\ee
is one of the first quantitative results of statistical physics.
The coefficient $\kb$ in front of the temperature is the Boltzmann constant.
The physical nature of the variables is irrelevant: $m$ can be the mass of a molecule and $v_x$ is the $x$-component of the velocity, $k$ can be elastic constant of a spring and $x$ can be the deformation, for a torsion magnetometer $x$ could be the angle, $\tilde I$ can be the moment of inertia of a molecule and $\omega$ is the angular velocity.
It is not a thoughtcrime to denote mass by $C$ and velocity by $U$.
Now the equipartition theorem is included in all high school textbooks.
Perhaps the most famous example is the mean kinetic energy of a single atom
\be
\frac12 \left < m \mathbf{v}^2 \right > = \frac32 \kb T, \quad
\mathbf{v}^2=v_x^2+v_y^2+v_z^2.
\ee
Here it is implicitly assumed that $\left < \mathbf{v} \right >=0$; no wind in the room.
Analogously, the mean voltage of the connected resistor and capacitor shown in Fig.~\ref{fig:cr} is zero $ \left <U \right >=0$.

In the beginning of the development of statistical physics
Albert Einstein suggested~\cite{Einstein:07} that Boltzmann constant $k_\mathrm{_B}$
can be determined using the equipartition theorem for and also time averaged thermally averaged square of the voltage $\left<U^2\right>$ of a capacitor $C$
\be
C \left <U^2\right> = k_\mathrm{_B} T.
\label{eqpart}
\ee
The first attempt for the realization of this idea was made by Habicht brothers~\cite{Habicht:10}
in 1910 but unfortunately their electrostatic amplifier with mechanically rotating parts suffered
from floating off the zero.
It is strange that for more than a century this simple experiment has not yet been performed.
Up to now, only Johnson and Schottky spectral noise experiments are conducted in
student laboratories.~\cite{Kraftmakher:95,Spiegel:95,MIT:13,Rodriguez:10,Wash:12}

Imagine that we have a set of capacitors with capacities given in Table~\ref{tbl:exp} with $V_2 \propto \left < U^2 \right >$.
\begin{center}
\begin{table}[h]
\begin{tabular}{ r  r }
		\hline
		&  \\ [-1em]
		$C$ [nF]  & $V_2$ [V]  \\ \tableline
			&  \\ [-1em]
			6.89 &  0.510 \\
			17.20 &  0.363 \\
			27.80 &  0.327 \\
			63.40 &  0.289 \\
			70.00 &  0.286 \\
			93.20 &  0.280 \\
			132.00 & 0.272 \\
\tableline
\end{tabular}
	\caption{Experimental results for different capacitors -- capacity and corresponding DC voltage measured by an ordinary cheap All-Sun DT-830B multimeter.}
	\label{tbl:exp}
\end{table}
\end{center}
According to the equipartition theorem at temperature 24$^\circ$C for the 6.89~nF and 132~nF we have for the voltage fluctuations $\delta U=\sqrt{\left<U^2\right>}=\sqrt{\kb T/C}$ correspondingly 771~nV and 176~nV.
If those voltages are amplified 1~million times, 771~mV and 176~mV can be measured even by cheap multimeters.
After several time constants $\tau_{_{RC}}=RC$,
 the random voltage $U(t)$ is already independent and in order to measure the mean square
\be
\left <U^2\right> = \frac{1}{\mathcal{T}} \int_0^\mathcal{T} U^2(t) \ud t,
\ee
it is necessary the averaging time $\mathcal{T}$ significantly to exceed $\tau_{_{RC}}$;
the accuracy is $\sqrt{\tau_{_{RC}}/\mathcal{T}}\ll 1$.

All those theorems are actually simple illustrations of the kinetic principle of the detailed balance applied to the most simple physical system -- the harmonic oscillator and can be considered as a comment to the Planck work~\cite{Planck:00} from the 19$^\mathrm{th}$ century.
In order to explain the spectral density of the black body radiation, Planck postulated the equidistant spectrum $E_n = \hbar \omega n$; it was the simplest realization of the Boltzmann idea for atomic structure of the energy.
In some sense, the discrete energy spectrum is associated with the quantum mechanics and it is interesting to trace the birthday of the quantum mechanics.
In the DPG conference in Halle 1891 Boltzmann presented his famous formula for the statistical interpretation of the entropy $S$ and the number of quantum states (in contemporary terminology) with one and the same energy.
Asked provocatively ``Do you consider that ...'' , unexpectedly for the auditorium, Boltzmann replied ``I do not exclude that energy can have atomic structure''.~\cite{atom}
In such a way, the birthyear of the quantum mechanics is 1891.

The manuscript by Waterston~\cite{Waterston} was rejected from publication by the polite referee report by Sir John William Lubbock: ``the paper is nothing but nonsense, ...''~\cite{nbn}
Even nowadays, the mathematical physicists feel obliged to convince the physicists that they did not lead Boltzmann to suicide.
In such a way, the Einstein idea to determine the Boltzmann constant by the equipartition theorem is in a good company.
The idea was not realised in the 20$^\mathrm{th}$ century and had to wait for the appearance of low-noise high frequency operational amplifiers to be realised in the students education.
The building of an experimental set-up for Boltzmann constant determination from scratch was a modus for the students in the University of Sofia to pass the exam in statistical physics.

The purpose of the present work is to present a simple self-made set-up
for determination of Boltzmann constant in high-school and university laboratories.
This is a good methodological illustration of the equipartition theorem.
First of all, the signal should be amplified million times $Y\approx 10^6$ using low-noise,
high-frequency operational amplifiers.
The problem of the floating off the zero is solved with large fast capacitors
sequentially connected to the gain resistors of the circuit depicted in Fig.~\ref{fig:circuit}.
Later on the signal has to be squared by an analog multiplier and
this squared signal has to be averaged by a low-pass filter.
Finally the time averaged signal $V$ is measured by an ordinary multimeter.
The simplicity of the experiment and set-up, as well as the availability of the electronic elements nowadays mean that this experimental set-up is suitable for high-school physics labs.
In such a way, our ultimate goal is to turn over the deteriorating trend in physics and engineering education worldwide, and this experimental set-up is a valuable tool in this endeavor.
In short, the set-up can be used by students to measure a fundamental constant $\kb$,
teachers can give explanations not only to physical principles but also to explain how the set-up works.
Enthusiastic teachers together with his or her students can make the set-up in one day.

\section{Theory of the experimental set-up}
\label{sec:theory}
\begin{figure*}[t]
\includegraphics[scale=0.25]{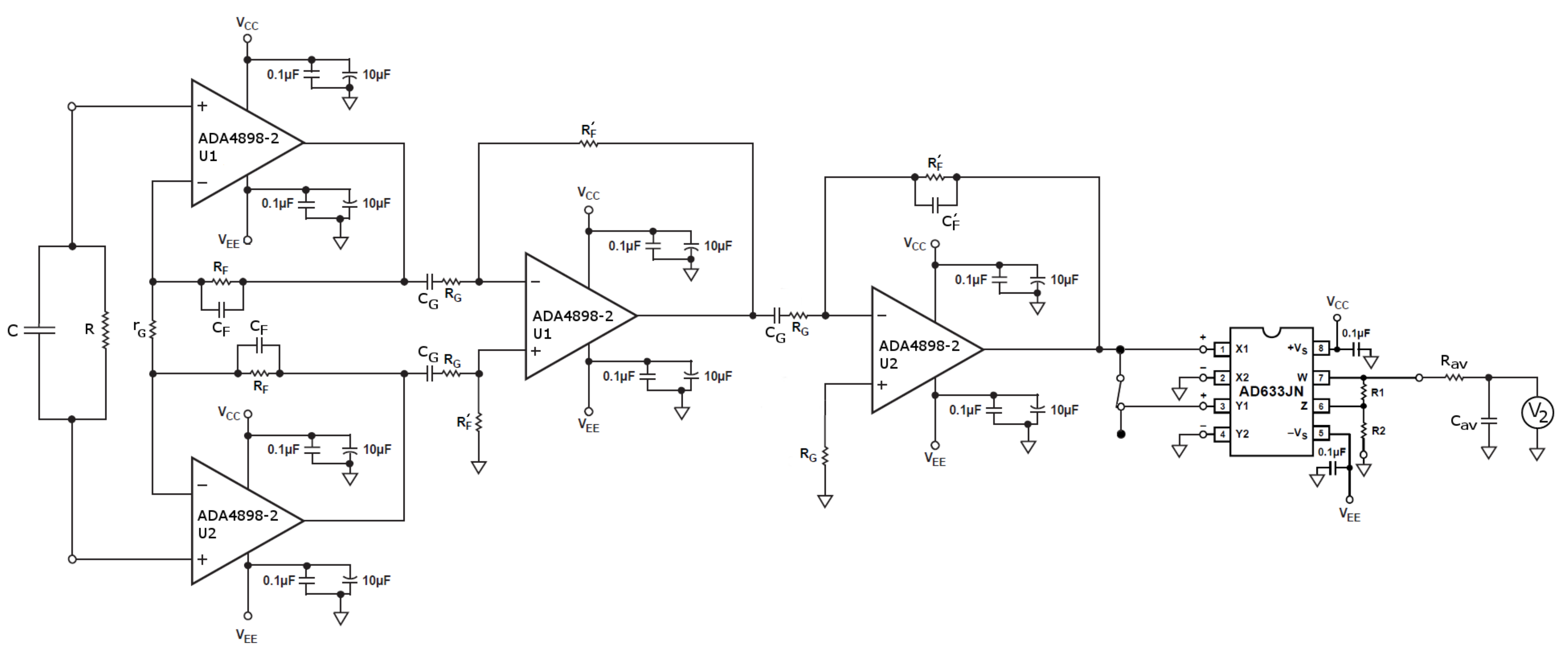}
\caption{A device for measurement of the voltage fluctuations $\delta U^2 = \left<(U - \left < U\right >)^2\right >= V_2/U^*$
of parallely connected capacitor $C$ and resistor $R$.
In the circuit we can recognize an instrumental amplifier,~\cite{ADA4817}
an inverting amplifier,~\cite{ADA4817} a multiplier with inputs connected in parallel,~\cite{AD633}
and an averaging low-frequency filter.
The expressions of the constants $y_1$ and $U^*$ by the circuit parameters is given in the text,
\Eqref{y1} and \Eqref{U*} correspondingly.
The voltage $V_2$ is measured with multimeters.
The circuit comprises two dual ADA4898-2~\cite{ADA4898} low-noise operational amplifiers.
The gain capacitors $C_\mathrm{G}$ are 10~$\mu$F MKS2 manufactured by WIMA.~\cite{MKS2}
In short, the voltmeter schematically represented in Fig.~\ref{fig:cr} is depicted in detail.
The amplified, squared and averaged voltage is measured by an ordinary multimeter.
As a whole the circuit can be described as a low-noise pre-amplifier followed by a true RMS-meter.
}
\label{fig:circuit}
\end{figure*}

As already stated in the previous section, the voltage fluctuations of nF capacitors in room temperature is of the order of hundred nV.
These nV should be amplified $10^6$ times and squared after that in order to be able to measure them with a relatively cheap and commercially available multimeters.
The circuit of the experimental set-up performing these actions is shown in Fig.~\ref{fig:circuit}.
The capacitor $C$, whose voltage fluctuations are to be measured is connected parallely to a resistor $R$.
This parallel circuit is connected to the two inputs of a buffer with amplification 
\be
y_1 =  1 + 2\frac{R_\mathrm{F}}{r_\mathrm{_G}},
\label{y1}
\ee
which is the first step of the amplifier.
Next a MKS2 WIMA type capacitor $C_\mathrm{G}$ is connected to each of the two outputs of the buffer to stop the voltage offset of the low-noise ADA4898-2 operational amplifiers.
The second step of the amplifier, which is an difference amplifier (DA) with amplification
\be
y_2 = - \frac{R_\mathrm{F}^\prime}{R_\mathrm{G}},
\label{y2}
\ee
is connected to both $C_\mathrm{G}$ capacitors, forming together with the buffer an instrumental amplifier.
The output of the DA is connected again to a MKS2 WIMA $C_\mathrm{G}$ capacitor with the same reason to stop the voltage offset and after $C_\mathrm{G}$ the third and last step, which is an inverting amplifier with amplification
\be
y_3 = - \frac{R_\mathrm{F}^\prime}{R_\mathrm{G}},
\label{y3}
\ee
is connected.
Finally, the total amplification of the amplifier 
\be
Y=y_1 y_2 y_3 =
\left(1+2\frac{R_\mathrm{F}}{r_\mathrm{_G}} \right )
\left(\frac{R_\mathrm{F}^\prime}{R_\mathrm{G}} \right)^{2}.
\label{Y}
\ee

The $Y$ times amplified signal is squared by an AD633 analog multiplier
by connecting the amplifier output to both inputs of the multiplier.
A voltage divider is connected after the amplifier according to the manufacturer instructions~\cite{AD633} and finally there is an averaging low pass filter consisting of a resistor $R_\mathrm{av}$ and a WIMA MKS2 type capacitor $C_\mathrm{av}$.
The values of all passive elements and voltage supplies for the ADA4898-2 operational amplifiers and AD633 analog multiplier are given in Table~\ref{tbl:values}.
\begin{center}
\begin{table}[h]
\begin{tabular}{ c  r }
		\hline
		&  \\ [-1em]
		Circuit element  & Value  \\ \tableline
			&  \\ [-1em]
			$R$ & 510~$\Omega$ \\
			$r_\mathrm{_G}$ & 20~$\Omega$ \\
			$R_\mathrm{F}$ &  1~k$\Omega$  \\
			$C_\mathrm{F}$ &  10~pF  \\ 
			$C_\mathrm{G}$ & 10~$\mu$F \\
			$R_\mathrm{G}$ &  100~$\Omega$  \\ 
			$R_\mathrm{F}^\prime$ & 10~k$\Omega$ \\
			$C_\mathrm{F}^\prime$ & 10~pF \\
			$R_1$ &  2~k$\Omega$  \\ 
			$R_2$ & 18~k$\Omega$  \\
			$R_\mathrm{av}$ & 1.5~M$\Omega$ \\
			$C_\mathrm{av}$ & 10~$\mu$F \\
			$R_\mathrm{V}$ & 1~M$\Omega$\cite{DT830B} \\
			$V_\mathrm{CC}$ & +9~V~\cite{ADA4898,AD633} \\
			$V_\mathrm{EE}$ & -9~V~\cite{ADA4898,AD633} \\
\tableline
\end{tabular}
	\caption{Table of the numerical values of the circuit elements from Fig.~\ref{fig:circuit}.}
	\label{tbl:values}
\end{table}
\end{center}
Lastly the resistor $R_\mathrm{av}$ forms another voltage divider with the internal resistance of the used DT-830B multimeter $R_\mathrm{V}$.\cite{DT830B}
For small values of $R_\mathrm{av}$ this voltage division is negligible, but in our case $R_\mathrm{av}$ and $R_\mathrm{V}$ are in the same order of magnitude and we have to take it into account in the final equation for the amplification.

The time dependent thermal fluctuations of the voltage $U(t)$ of the parallely connected $C$ and $R$ is amplified $Y$ times
\be
U_\mathrm{amp}=Y U.
\label{ampl}
\ee
Then the amplified voltage $U_\mathrm{amp}$ is squared by an analog multiplier.
We use the circuit depicted in Ref.~\onlinecite[Fig.~17]{AD633}, 
where X2~=~Y2~=~0, 
X1~=~Y1~=~$U_\mathrm{amp}(t)$
and W~$\equiv U_\square$, i.e. we have
\be
U_\square = \frac{U_\mathrm{amp}^2}{U_\mathrm{m}} \frac{R_1+R_2}{R_1},
\label{square}
\ee
where the constant factor for this multiplies AD633~\cite{AD633} 
$U_\mathrm{m}=10$~V.
Actually we have a voltage divider for which Z~=~W$R_2/(R_1+R_2)$ and S~=~0.
The squared voltage is averaged by an averaging low pass filter with large time constant 
$\tau_\mathrm{av}=R_\mathrm{av}C_\mathrm{av}$ of order of a quarter a minute.
The large $R_\mathrm{av}$resistance is comparable with the internal resistance of the voltmeter 
$R_\mathrm{V}$ and for time averaged DC voltage shown by the voltmeter we have to take into account a second voltage divider
\be
V_2 = \left<U_\square\right> \frac{R_\mathrm{V}}{R_\mathrm{V}+R_\mathrm{av}},
\label{avrg}
\ee
while $\left < V_1 (t) \right >=0$ the fluctuations of the electric noise create non-zero square.
Combined together equation for amplifying  Eq.~(\ref{ampl}), 
squaring Eq.~(\ref{square})
and averaging Eq.~(\ref{avrg}) we arrive at  the transfer function of the amplifier 
\be
V_2 = \frac{Y^2}{U_\mathrm{m}}  \frac{R_1+R_2}{R_1}
\frac{R_\mathrm{V}}{R_\mathrm{V}+R_\mathrm{av}} \left<U^2(t)\right>,
\label{eq:V2}
\ee
where brackets mean time averaging at steady spectral density of the noise.
This can also be written in the form
\be
\left <U^2\right> = U^* V_2 \label{U2},
\ee
where the time averaged square of the investigated voltage $\left <U^2\right>$ and the 
constant voltage measured by a voltmeter $V$ is related by a constant with dimension of voltage 
\be
\frac1{U^*} \equiv\frac{Y^2}{U_\mathrm{m}} \frac{R_1+R_2}{R_1}
\frac{R_\mathrm{V}}{R_\mathrm{V}+R_\mathrm{av}}.
\label{U*}
\ee
If we substitute here the expressions for $Y$ from \Eqref{Y}, \Eqref{y1}, \Eqref{y2} and \Eqref{y3}
we obtain the final expression for the voltage constant describing our set-up
\be
\frac1{U^*} =\frac{1}{U_\mathrm{m}}\left(1\!+\!2\frac{R_\mathrm{F}}{r_\mathrm{_G}} \right )^{\!2}
\left(\frac{R_\mathrm{F}^\prime}{R_\mathrm{G}} \right)^{\!4} 
\frac{R_1\!+\!R_2}{R_1}
\frac{R_\mathrm{V}}{R_\mathrm{V}\!+\!R_\mathrm{av}}.
\label{U*}
\ee
As a whole, the circuit can be characterized as a low noise pre-amplifier (instrumental amplifier followed by an inverting amplifier all of them based on ADA4898-2\cite{ADA4898}) and a sequential true RMS-meter.
If necessary, the cheap AD633\cite{AD633} (maximum total error 2\% of full scale, small signal bandwidth 1~MHz) multiplier can be substituted with more precise AD835\cite{AD835} (total error 0.1\% of full scale, small signal bandwidth 250~MHz). 
But even in this case, the combination of a commercial low-noise pre-amplifier and a true RMS-meter costs one or even two orders of magnitude more than our experimental set-up that can be easily reproduced in any school physics classroom.

The even more detailed derivation of all those formulae requires only Ohm law applied to a voltage divider and a sequential chain of simple problems for high-school students is described in great detail
in Ref.~\onlinecite{EPO5} and it is remarkable that even a high-school student was able to solve the university electronic problem.

The substitution of $ \left <U^2\right>$ from the equipartition theorem \Eqref{eqpart} into 
the property of the circuit we derived \Eqref{U2}
gives a linear dependence between the DC voltage $V_2$ and the reciprocal capacitance $1/C$
\be
V_2 = q_0 \frac1{C} + v_0, \quad q_0\equiv \frac{\mathcal{Z}k_\mathrm{_B}T}{U^*}, \quad \mathcal{Z}\approx1.
\label{q0}
\ee
The irrelevant for the experiment constant $v_0$ describes the internal noise of the circuit determined mainly by the voltage noise of the first dual ADA4898-2 amplifier in the buffer (double non-inverting amplifier with a virtual common point).
In order $v_0$ to be small, we need to use low-noise operational amplifiers.

The slope $q_0$ of the linear regression $V_2$ versus $1/C$
\be
q_0\equiv\left.\frac{\mathrm{d} V_2}{\mathrm{d} C^{-1}}\right|_\mathrm{regr}
\ee
determines the Boltzmann constant 
\be
k_\mathrm{_B}=\frac{q_0U^*}{\mathcal{Z}T}.
\label{kb}
\ee

For high-school students the correction multiplier 
$\mathcal{Z}=1-\varepsilon$ is indistinguishable from one,
but for university students at undergraduate level the several percent correction $\varepsilon$
can be calculated analyzing the frequency dependence of the amplifier as it is described in great detail in Appendix~\ref{Electronics_in_nutshell}.

\section{Frequency dependent considerations}
\label{Sec:Freq}

Up to now, the frequency dependence have been neglected.
For high school and even undergraduate level this is perfectly acceptable but for a detailed analysis of the amplifier the frequency dependence has to be included.
This analysis is far more complicated and is performed in Sec.~\ref{Electronics_in_nutshell}.

In this section we present a short analysis of time constants and frequencies of the presented experimental set-up.

The ADA4898-2 operational amplifier time constant $\tau$ is calculated from the crossover frequency $f_0$ with $\tau= 1/2\pi f_0$.\cite{ADA4898}
For amplification maximally close to the frequency independent one (the considered scenario in the last section), $\tau \ll \tau_\mathrm{_F}$, where $\tau_\mathrm{_F}=R_{_\mathrm F}C_{_\mathrm F}$ is the time constant of the buffer feedback.
In other words, the operational amplifier should not ``feel'' that there is a capacitance in its feedback.
The amplified by the buffer signal now has a higher time constant $\tau_{_\mathrm A}=\tau y_1$ or $f_{_\mathrm A}=f_0/y_1$ or lower frequency due to the lower amplification of higher frequencies by the operational amplifier and $\tau_{_\mathrm{A}} \gg \tau_\mathrm{_F}$ because of the requirement for the feedback capacitance.
After the buffer this signal passes through a high-pass filter consisting of the resistor $R_{_\mathrm G}$ and the capacitor $C_{_\mathrm G}$ with a large time constant $\tau_\mathrm{_G}=R_{_\mathrm G}C_{_\mathrm G}$, whose function is to filter out the voltage offsets of the ADA4898-2 operational amplifiers of the buffer.
For a proper operation of the amplifier up to now, the time constants should be ordered
\be
\tau \ll \tau_\mathrm{_F} \ll \tau_{_\mathrm A} \ll \tau_{_\mathrm G}.
\ee

There is another capacitor $C_{_\mathrm F}^\prime$ in the feedback of the IA and analogously to the buffer, its presence should be barely ``felt'', therefore the time constant of the IA feedback $\tau_\mathrm{_F}^\prime=R_{_\mathrm F}^\prime C_{_\mathrm F}^\prime < \tau_\mathrm{_A}$.
And finally the averaging low pass filter should have the largest time constant $\tau_\mathrm{av}=R_{\mathrm {av}} C_{\mathrm {av}}$ in order to reliably average the squared voltage.

Now let us return to the input signal, which is the time dependent thermal fluctuations of the voltage of the parallely connected capacitor $C$ and resistor $R$, whose time constant is $\tau_{_{RC}}=RC$.
This time constant should be much larger than $\tau_\mathrm{_A}$ for a maximum amplification but much smaller than $\tau_{_\mathrm G}$ for a maximum transfer between the three steps of the whole amplifier.
Therefore, the requirement for the optimal  amplification of our experimental set-up is 
\be
\tau \ll \tau_\mathrm{_F} \ll \tau_\mathrm{_F}^\prime < \tau_{_\mathrm A} \ll \tau_{_{RC}} \ll \tau_{_\mathrm G} \ll \tau_\mathrm{av}.
\ee
A set of the calculated time constants, frequencies and additional calculated parameters of the experimental set-up is given in Table~\ref{tbl:pars},
\begin{center}
\begin{table}[h]
\begin{tabular}{ c  r }
		\hline
		&  \\ [-1em]
		Calculated parameter  & Value  \\
\tableline
			&  \\ [-1em]
			$f_0$ & 65~MHz~\cite{ADA4898} \\
			$\tau$ & 3.94~ns\\
			$\tau_\mathrm{_A}$ & 398~ns \\
			$f_\mathrm{_A}$ & 396~kHz \\
			$\tau_{_{RC}}$ & (7.65--66.3)~$\mu$s\\
			$\tau_\mathrm{_F} $ & 10~ns \\
			$\tau_\mathrm{_G}$ & 1~ms \\
			$f_\mathrm{_G}$ & 159~Hz \\
			$\tau_\mathrm{_F}^\prime$ & 100~ns \\
			$\tau_\mathrm{av}$ & 15~s \\
			$Y$ & $1.01 \times 10^6$ \\
			$U_\mathrm{m}$ &  10~V~\cite{AD633} \\
			$U^*$ & 2.45~pV \\
\tableline
\end{tabular}
	\caption{Table of the calculated parameters necessary for the analysis of the circuit
	(time constants, frequencies, voltages and linear amplification).
$f_0$ is the -3dB bandwidth of the ADA4898, $\tau \equiv 1/2\pi f_0,$
$f_{_\mathrm A}=f_0/y_1,$ $\tau_{_\mathrm A}=\tau y_1,$
$\tau_{_{RC}}=RC,$ $\tau_\mathrm{_F}=R_{_\mathrm F}C_{_\mathrm F},$
$\tau_\mathrm{_F}^\prime=R_{_\mathrm F}^\prime C_{_\mathrm F}^\prime,$
$\tau_\mathrm{_G}=R_{_\mathrm G}C_{_\mathrm G},$
$\tau_\mathrm{av}=R_{\mathrm {av}} C_{\mathrm {av}},$
and \\
$\tau \ll \tau_\mathrm{_F}, \tau_\mathrm{_F}^\prime, \tau_{_\mathrm A}
\ll \tau_{_{RC}}\ll \tau_{_\mathrm G}\ll \tau_\mathrm{av}.$
	           }
	\label{tbl:pars}
\end{table}
\end{center}

Why do we need capacitors?
1) Even during the first attempt in measuring the Boltzmann constant, the Habicht brothers noted that their amplifier suffers from floating off the zero.~\cite{Einstein:07,Habicht:10}
Even nowadays, there are no low-noise auto-zero operational amplifiers commercially available.
In order to remove this unpleasant floating off the zero and the low-noise 1/f noise in frequencies over 100~Hz, large $C_\mathrm{G}=10~\mu$F metallized polyester capacitors, whose price is comparable to that of the batteries and operational amplifiers, are put on the path of the signal.
2) The small ceramic $C_\mathrm{F}$ and $C_\mathrm{F}^\prime$ connected parallely to the feedback resistors reduce the amplification in high frequencies ($> 1/R_\mathrm{F} C_\mathrm{F}$, $>1/R_\mathrm{F}^\prime C_\mathrm{F}^\prime$) and in this way stabilise the amplifies in terms of ringing.

In short, the removal of the offset and the ringing of the amplifier causes the addition of capacitors.
If we additionally assume the operational amplifiers to be ideal $\tau=0$, in the approximation
$C_\mathrm{F}=C_\mathrm{F}^\prime=0$ and $C_\mathrm{G}=\infty$,
for the amplification of the amplifier we have the frequency independent approximation
$Y(\omega) \approx y_1 y_2 y_3$, for which we have put lots of effort to be a very good approximation in engineering the amplifier. 
Infinite capacitance  $C_\mathrm{G}=\infty$ means short circuit and signal is going directly to the gain resistor, while $C_\mathrm{F}=C_\mathrm{F}^\prime=0$ means absence of the parallel connected capacitors. 

\section{Experiment}
\label{sec:experiment}
\begin{figure}[h]
\includegraphics[scale=0.13]{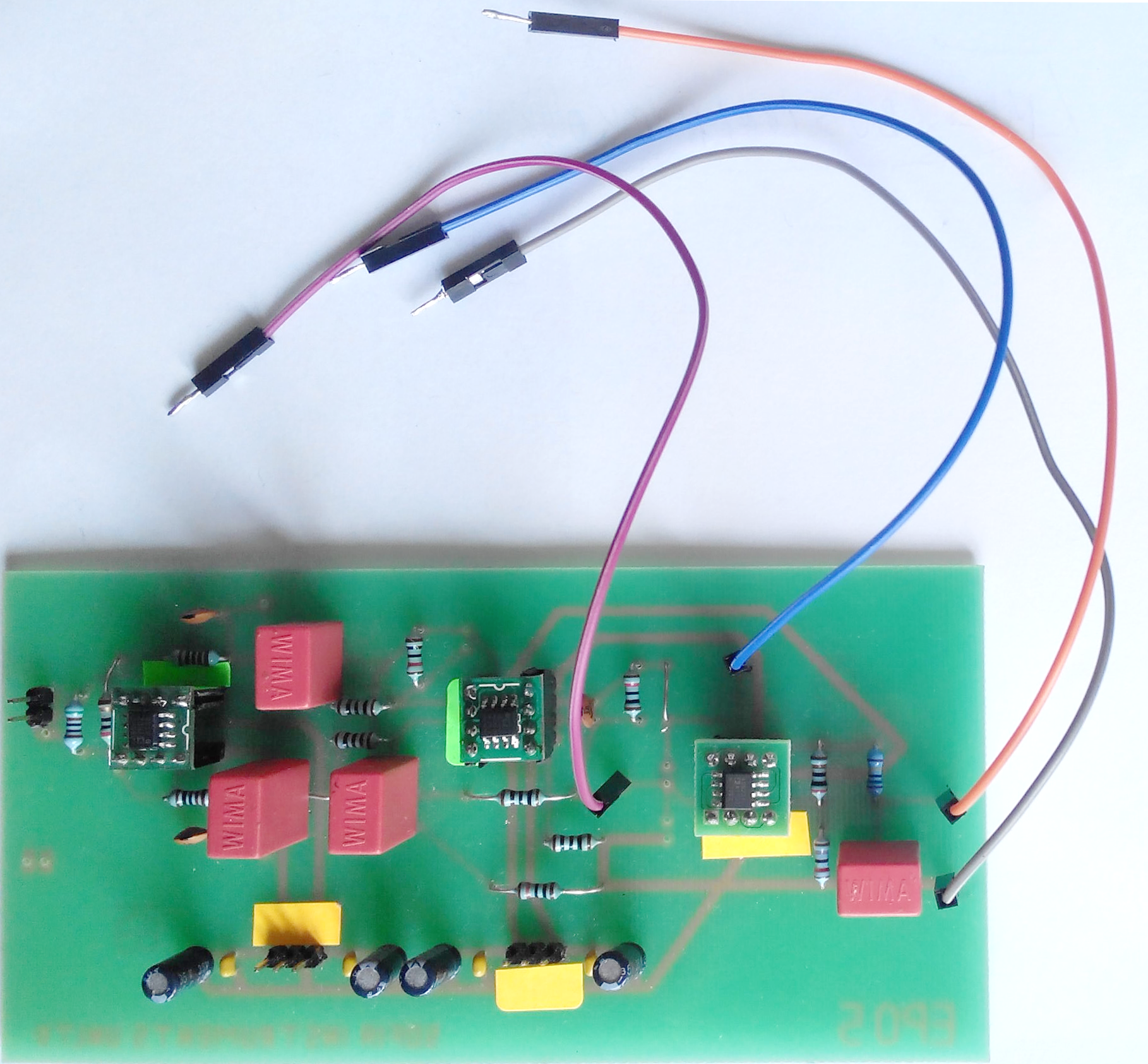}
\caption{A realization of the circuit on a printable circuit board given in EPO5.~\cite{EPO5}}
\label{fig:epo5}
\end{figure}

\begin{figure}[h]
\includegraphics[scale=0.55]{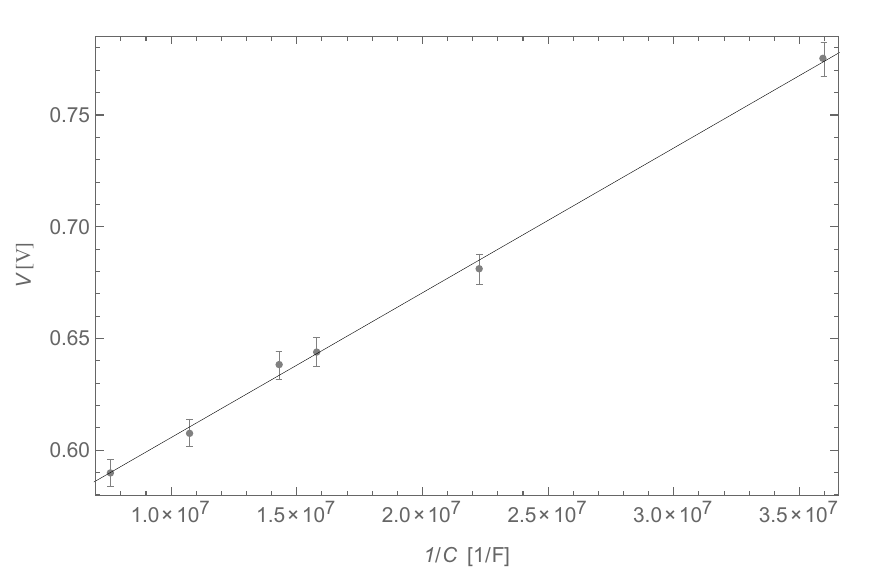}
\caption{Correlation between the voltage $V$ measured by the right-hand side multimeter
from Fig.~\ref{fig:circuit} and the reciprocal capacitance of the capacitor $C$ at the
beginning of the same circuit.
The linear regression with high correlation coefficient $\rho=0.9997$ is a consequence of the
equipartition theorem \Eqref{eqpart}; see also \Eqref{U2}.
The slope of the straight line $q_0$ determines the Boltzmann constant \Eqref{kb}.}
\label{fig:regr}
\end{figure}

A photograph of the realized circuit on a PCB is given in Fig.~\ref{fig:epo5},
which was produced in 200 copies and given to high school students in the 5$^\mathrm{th}$ Experimental Physics Olympiad~\cite{EPO5} (EPO5) is shown in Fig.~\ref{fig:epo5}.
The set of the parameters from the circuit and the integral schemes is listed in Table~\ref{tbl:values}.
The experimental data measured with an ordinary cheap All-Sun DT-830B multimeter corresponding to \Eqref{kb} is presented in Table~\ref{tbl:exp} and graphically shown in Fig.~\ref{fig:regr}.
The obtained value of the slope $q_0=(1718 \pm 18)$~pC and
the achieved accuracy for determination of Boltzmann constant is rather good
\be
k_\mathrm{_B} = (1.40 \pm 0.08) \times 10^{-23}~\mathrm{J/K}
\ee
for a \$50 set-up which can be further elaborated.
The random thermal voltage can be observed with an oscilloscope,
the temperature $T$ of the resistor $R$ can be varied from the freezing to boiling point of the water  
but we are presenting only the simplest experiment with varying only the capacitor in room temperature which can be realized in every high school.
It is intriguing to measure a fundamental constant by a set-up which can be created 
from scratch within a week by a novice. 

\section{Conclusions}
\label{sec:conclusion}

Some administrators related to education in the third world countries can conclude that
very few high schools or introductory college courses would have the background,
interest and/or resources to adopt this experiment.
However, the equipartition theorem is already in high school education for more than a century.
More than a century it is taught that the temperature is related with motion and fluctuations.
The vacuum technique will remain expensive and even in good universities Maxwell velocity distribution 
is not experimentally demonstrated.
However, for the last century electronics made a significant progress and prices are reduced thousand times.
That is why the described set-up can be industrially produced for illustration 
of the thermal fluctuations in high schools even in the framework of the existing school programs. 
But even without commercially available set-ups
the experimental set-up contains 3~integral circuits: 2~operational amplifiers,
a multiplier, 3~large fast capacitors and 9~V batteries.
Such a set-up can be performed within one day in every high-school and complete the teaching of
thermal fluctuations and electronics.
The problem of Boltzmann constant measurement with the described experimental setup was given in EPO5~\cite{EPO5} to more than a hundred high school students and more than 10\% of them were able to perform the experiment and to process their data to obtain a final answer
without any special training only in the framework of the standard education.
We can certainly conclude that the whole experiment is appropriate for high-school physics laboratories all over the world, where in addition a hardware or software oscilloscope can be used to visualize the 1 million times amplified thermal noise.
As a by product of EPO5 the authors of the present article can send free of charge 
to the physics laboratory
one set-up to the first 137 teachers which will write us.
When the set-up PCB version is outsourced in China,
the measurement of Boltzmann constant can reach any student from the
1-st, 2-nd and perhaps 3-rd world before WWIII.
In short, we have arrived to the conclusion that it is time to introduce in the
high school education in physics simple experimental set-ups for determination of fundamental constants  not only $\kb$, but also electron charge $q_e$~\cite{Electron}, speed of light $c$~\cite{Light} and Planck's constant $\hbar$~\cite{Planck} as well.
The touching to the fundamental determines the thinking of the next generation. 


\acknowledgments{}

The authors are grateful to Vasil Yordanov for his contribution at the early stages of the present
research,~\cite{MYV} to Alexander Petkov for making the first measurements, to Gary White for stimulating comments,  to Nikolay Zografov for introducing order in the lab, to Andreana Andreeva for animation of the spirit in the lab, to Riste Popeski-Dimovski, Marina Poposka, Sladjana Nikolic, Slavoljub Mitic and Stojan Manolev for the invaluable help and unforgettable moments during EPO.



\appendix

\section{Analog electronics in a nutshell}
\label{Electronics_in_nutshell}

This appendix is addressed to colleagues involved with construction of similar devices and modifying the scheme.
This recall of the standard electronics notion is not addressed to the students.

\subsection{Operational amplifier master equation}
In the beginning was the approximate master equation of
operational amplifiers
\begin{equation}
\label{MasterOpAmp}
\hat{G}^{-1}U_0(t)=\left(\frac1{G_0}+\tau\frac{\mathrm{d}}{\mathrm{d}t}\right)U_0=U_+-U_-,
\end{equation}
giving the relation between the output voltage $U_0(t)$ and
difference of (+) and (-) inputs of operational amplifier, cf. Ref.~\onlinecite{Lee:12}.
For harmonic signals, introducing j-imaginary unit
\begin{equation}
U\propto\mathrm{e}^{\,\mathrm{j}\omega t}
=\mathrm{e}^{\,s t},\quad \mathrm{j}=-\mathrm{i},\quad
s\equiv\mathrm{j}\omega,
\end{equation}
we have
\begin{eqnarray}
\label{MasterFourier}&&
G^{-1}(\omega)U_0=\left(\frac1{G_0}+\mathrm{j}\omega\tau\right)U_0=U_+-U_-,
\\&&
G^{-1}\!=\!\left(\frac1{G_0}\!+\!s\tau\right)\!\frac{1+a_1s+a_2s^2+a_3s^3+\dots}
{1+b_1s+b_2s^2+b_3s^3+\dots},\end{eqnarray}
where in the second row we present a Pad\'e approximant for the
frequency dependent open loop gain.
For the used by us operational amplifier ADA4898\cite{ADA4898}
the static open loop gain is approximately 100~dB
\begin{equation}
G_0=10^5,\qquad
f_0\equiv\frac1{2\pi\tau}=65\;\mathrm{MHz}
\end{equation}
and the time $\tau$ constant is parameterized by the crossover frequency $f_0$.

In the next subsections we will recall how the frequency dependent open loop gain
determines the frequency dependent transmission function of different amplifiers.

\subsection{Buffer}
\begin{figure}[h]
\includegraphics[scale=0.25]{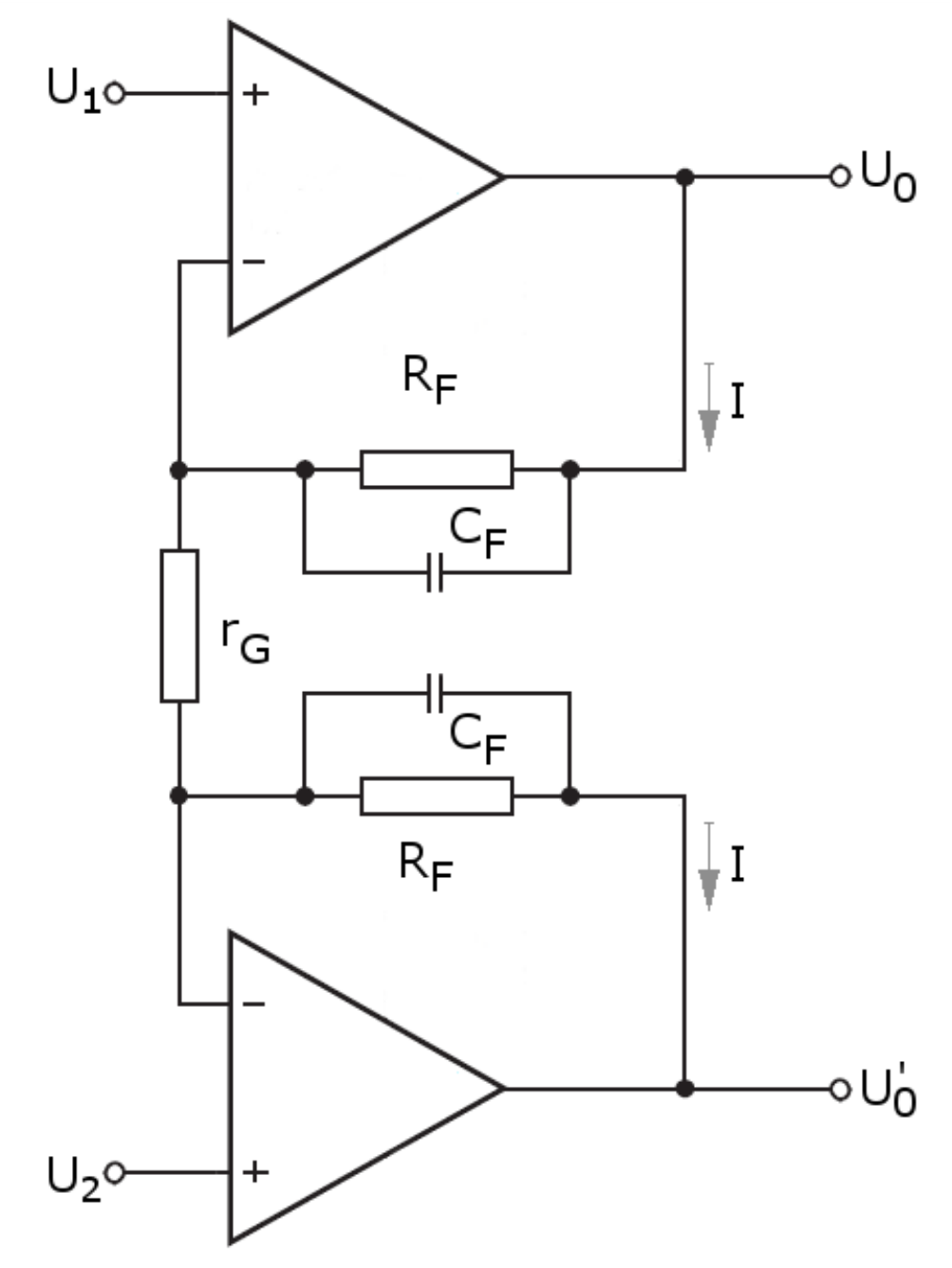}
\caption{Circuit of a buffer with capacitance in the feedback.} 
\label{fig:buffer}
\end{figure}

For the buffer, consisting of two non-inverting amplifiers (NIA) depicted in Fig.~\ref{fig:buffer},
the input voltages $U_1$ and $U_2$ are applied directly to the (+) inputs of each OpAmp 
$U_+=U_1$, and $U_+^\prime=U_2$, here primed notations refer to the lower NIA.
Since no currents enter in the OpAmp, the only current $I$ starts from the output $U_0$, passes through the upper feedback $Z_\mathrm{F}$, gain resistor $r_\mathrm{_G}$, lower feedback $Z_\mathrm{F}$ and terminates in the output $U_0^\prime$.
From this closed loop, it is straightforward to obtain an expression for the current 
\be
I=\frac{(U_0^\prime-U_0)\equiv -\Delta U_0}{2 Z_\mathrm{F} + r_\mathrm{_G}},
\quad \frac1{Z_\mathrm{F}(\omega)}=\frac1{R_\mathrm{F}}+\mathrm{j}\omega C_\mathrm{F}.
\ee
The voltage difference between the (-) inputs of both OpAmp of the buffer is
\be
U_- - U_-^\prime = -I r_\mathrm{_G} 
= \Delta U_0 \frac{ r_\mathrm{_G}}{2 Z_\mathrm{F} + r_\mathrm{_G}}.
\label{eq:nia_u-}
\ee
The master equations for both operational amplifiers are
\begin{eqnarray}
&& U_0 G^{-1}(\omega)=U_1-U_- \\
&& U_0^\prime G^{-1}(\omega)=U_2-U_-^\prime, 
\end{eqnarray}
and after subtracting them, we obtain a single equation
\begin{equation}
\Delta U_0 G^{-1} = (U_1-U_2) - (U_--U_-^\prime).
\end{equation}
Substituting \Eqref{eq:nia_u-} into the last equation, the frequency dependent amplification of the buffer
\bear
&& \Upsilon_\mathrm{NIA}(\omega) \equiv \frac{\Delta U_0}{U_1-U_2} = 
\frac1{G^{-1}(\omega)+y^{-1}(\omega)}, \label{aNIA} \\
&& y(\omega) \equiv \frac{Z_\mathrm{F}(\omega)}{r_\mathrm{_G}}+1
\eear
in agreement with Ref.~\onlinecite[Eq.~(4)]{ADA4817}.
Using complex numbers in programming we can simply calculate 
$|\Upsilon_\mathrm{NIA}(\omega)|^2$, however using only real numbers we
have to apply some efforts in complex algebra and
after a straightforward calculation from \ref{aNIA} we obtain
\begin{eqnarray}&&
\left|\Upsilon_\mathrm{NIA}(\omega)\right|^2\\&&\nonumber=\frac{\mathcal{N}^2(\omega)}
{\left[G_0^{-1}\mathcal{N}+ (y_1^{-1}+\omega^2\tau_s^2)\right]^2+(\omega\tau_s)^2\left[\dfrac{\tau}{\tau_s}\mathcal{N}+I_0\right]^2},
\end{eqnarray}
where
\begin{equation}
\mathcal{N}(\omega)=1+\omega^2\tau_s^2,\quad
I_0\equiv 1-y_1^{-1},
\quad
\tau_s\equiv \frac{C_\mathrm{F} R_\mathrm{F}}{y_1}.
\end{equation}
The calculation of the pass-bandwidth requires the square of the modulus of the complex amplifications for all steps of the amplifier which we are presenting in full detail.  

\subsection{Difference Amplifier}

\begin{figure}[h]
\includegraphics[scale=0.25]{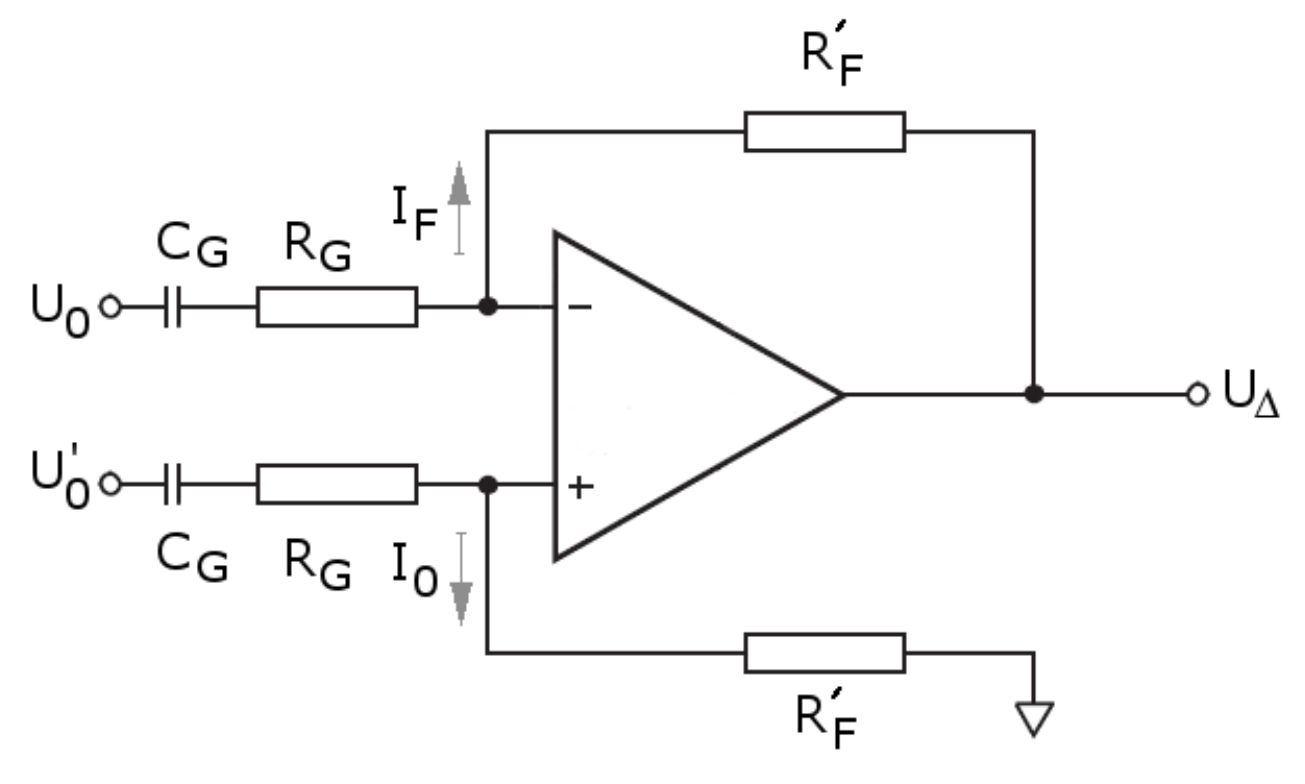}
\caption{Circuit of a difference amplifier with capacitance in the input.} 
\label{fig:da}
\end{figure}

The difference amplifier shown in Fig.~\ref{fig:da} has a capacitor $C_\mathrm{G}$ at each of its inputs.
The current $I_\mathrm{F}$ flows from the first input $U_0$ of the difference amplifier to the output $U_\Delta$ and
\be
I_\mathrm{F}=\frac{U_\Delta-U_0}{Z_\mathrm{G}+R_\mathrm{F}^\prime}, \qquad
Z_\mathrm{G}(\omega)=R_\mathrm{G}+\frac1{\mathrm{j} \omega C_\mathrm{G}},
\ee
while the current $I_0$ flows from the second input $U_0^\prime$ to the common point and
\be
I_\mathrm{0}=\frac{-U_0^\prime}{Z_\mathrm{G}+R_\mathrm{F}^\prime}.
\ee
The voltage drops at the inputs of the OpAmp
\bear
&& U_-=I_\mathrm{F} Z_\mathrm{G}+U_0=
\frac{U_\Delta-U_0}{Z_\mathrm{G}+R_\mathrm{F}^\prime} Z_\mathrm{G}+U_0, \\
&& U_+=I_0 Z_\mathrm{G}+U_0^\prime=
\frac{-U_0^\prime}{Z_\mathrm{G}+R_\mathrm{F}^\prime} Z_\mathrm{G}+U_0^\prime.
\eear
Subtracting the last two equations, we obtain an expression for the voltage difference of the OpAmp inputs
\be
U_+-U_-= -(U_0-U_0^\prime) \!\! 
\left ( 1 - \frac{Z_\mathrm{G}}{Z_\mathrm{G}\!+\!R_\mathrm{F}^\prime} \right )
-\frac{U_\Delta Z_\mathrm{G}}{Z_\mathrm{G}\!+\!R_\mathrm{F}^\prime}.
\ee
This expression is equal to $G^{-1}(\omega) U_\Delta$ according to \Eqref{MasterOpAmp} (here $U_\Delta$ is the output voltage) and therefore after a little rearrangement of the terms
\be
U_\Delta \left ( G^{-1} + \frac{Z_\mathrm{G}}{Z_\mathrm{G}+R_\mathrm{F}^\prime} \right ) =
-(U_0-U_0^\prime) \frac{R_\mathrm{F}^\prime}{Z_\mathrm{G}+R_\mathrm{F}^\prime}.
\ee
The frequency dependent amplification of the difference amplifier therefore is
\be
\Upsilon_\Delta (\omega) \equiv \frac{U_\Delta}{U_0-U_0^\prime} =
\frac{-1}{\Lambda(\omega)+G^{-1}(\omega)[1+\Lambda(\omega)]},
\label{aD} 
\ee
where $\Lambda(\omega) \equiv Z_\mathrm{G}(\omega)/R_\mathrm{F}^\prime$.
The calculation of the pass-bandwidth requires the square of the modulus of the complex amplification, which after a straightforward calculation from \ref{aD}
\begin{eqnarray}
&&
\left|\Upsilon_\Delta(\omega)\right|^2 \label{absD}
\\&&\nonumber=\frac{(\omega\tau_\mathrm{g})^2}
{\left[1\!+\!G_0^{-1}\!-\!M\omega^2\tau\tau_\mathrm{g}\right]^2
+(\omega\tau_\mathrm{g})^2\left[\dfrac{\tau}{\tau_\mathrm{g}}\!+\!\Lambda_0\!+\!G_0^{-1}M\right]^2}, \\ 
&&\nonumber\quad 
\Lambda_0\equiv\frac{R_\mathrm{G}}{R_\mathrm{F}^\prime},\qquad M\equiv 1+\Lambda_0,\qquad
\tau_\mathrm{g} \equiv C_\mathrm{G} R_\mathrm{F}^\prime.
\end{eqnarray}
This complicated expression is necessary if we wish to use real numbers in the numerical integration necessary to be preformed for the pass bandwidth. 
The last step of the amplifier is the inverting one described in the next subsection.

\subsection{Inverting Amplifier}

Setting $U_0^\prime=0$ in the difference amplifier from the last subsection, we obtain an inverting amplifier (IA).
In our case there is one more difference, which is a capacitor $C_\mathrm{F}$ connected parallely to the feedback resistor $R_\mathrm{F}^\prime$ and therefore the reciprocal of the impedance feedback
\be
\frac1{Z_\mathrm{F}^\prime(\omega)}=\frac1{R_\mathrm{F}^\prime}+\mathrm{j}\omega C_\mathrm{F}.
\ee
Re-denoting $\Lambda(\omega)$ to $\Gamma(\omega) \equiv Z_\mathrm{G}(\omega)/Z_\mathrm{F}^\prime$, the frequency dependent amplification of the IA
\be
\Upsilon_\mathrm{IA} (\omega) =
-\frac{1}{\Gamma(\omega)+G^{-1}(\omega)[1+\Gamma(\omega)]}
\label{aIA} 
\ee
in agreement with Ref.~\onlinecite[Eq.~(7)]{ADA4817}.
This agreement gives an implicit proof of the applicability of the time-dependent Eq.~\ref{MasterOpAmp} and its frequency transformation Eq.~\ref{MasterFourier} for circuits with low-noise operational amplifiers ADA4898, ADA4817, AD711 etc.
The calculation of the pass-bandwidth requires the square of the modulus of the complex amplification, which after a straightforward but a bit more complicated calculation from \ref{aIA}
\begin{widetext}
\be
\left|\Upsilon_\mathrm{IA}(\omega)\right|^2=\frac{(\omega\tau_\mathrm{g})^2}
{\left[(1\!+\!G_0^{-1})(1\!-\!\omega^2\tau_\mathrm{fg}\tau_\mathrm{g})\!-\!M\omega^2\tau\tau_\mathrm{g}\!-\!\omega^2 \tau \tau_\mathrm{f}^\prime\right]^2
+(\omega\tau_\mathrm{g})^2\left[\dfrac{\tau}{\tau_\mathrm{g}}\!+\!\Lambda_0\!+\!G_0^{-1}M\!+\!\dfrac{\tau_\mathrm{f}^\prime}{\tau_\mathrm{g}}(G_0^{-1}\!+\!1)\!-\!\omega^2\tau \tau_\mathrm{fg} \right]^2},
\label{absIA}
\ee
\end{widetext}
where 
\be
\tau_\mathrm{f}^\prime \equiv C_\mathrm{F} R_\mathrm{F}^\prime, \qquad
\tau_\mathrm{fg} \equiv C_\mathrm{F} R_\mathrm{G}.
\ee
For $C_\mathrm{F}=0$, $\tau_\mathrm{f}^\prime=\tau_\mathrm{fg}=0$ and the modulus of the  amplification of the IA \Eqref{absIA} becomes equal to the modulus of the amplification of the difference amplifier \Eqref{absD}.

Finally, for the whole amplifier we have
\begin{equation}
\Upsilon (\omega)=\Upsilon_\mathrm{NIA}(\omega) \Upsilon_\Delta(\omega) \Upsilon_\mathrm{IA}(\omega).
\label{Ups}
\end{equation}

\subsection{Application of the Nyquist theorem for our circuit}

In order to calculate the mean square of the amplified signal
$\left < U_\mathrm{amp}^2 (t)\right >$
we have to apply the Nyquist theorem for the spectral density of the thermal noise of the parallely connected capacitor $C$ and resistor $R$ at the input of the circuit.
According this theorem the spectral density of the noise is given by the real part of the impedance
\begin{align}
(U^2)_f = 4 \kb T R(\omega), \quad R(\omega) = \Re(Z(\omega)), 
\quad \hbar\omega\ll \kb T. 
\end{align}
A new pedagogical re-derivation of Nyquist theorem and its extensions is given in
Appendix~\ref{Apendix_theory}.

Although the quantum physics was created from the Planck explanation of the spectral density of the electromagnetic fluctuations and 3 Noble prizes were given for the black body radiation, there is no definite notation for the spectral density of the voltage noise in electronics.
For instance, in Ref.~\onlinecite[Chap.~4]{Kittel:80}, Ref.~\onlinecite[Eq.~78.3]{LL9}, where
$(\mathcal{E}^2)_\omega = (\mathcal{E}^2)_f/2$ or Ref.~\onlinecite[Chap.~24]{McCombie:71}, where $G_V = (\mathcal{E}^2)_f/2 \pi$, or Ref.~\onlinecite[Eq.~89.3]{LL8}, where 
$(\mathcal{E}^2)_\omega = (\mathcal{E}^2)_f/ 4 \pi $, or Ref.~\onlinecite[Eq.~(1.4.47)]{Gardiner:85}, where $S(\omega) = (\mathcal{E}^2)_f/ 4 \pi $.

\begin{figure}[h]
\includegraphics[scale=0.55]{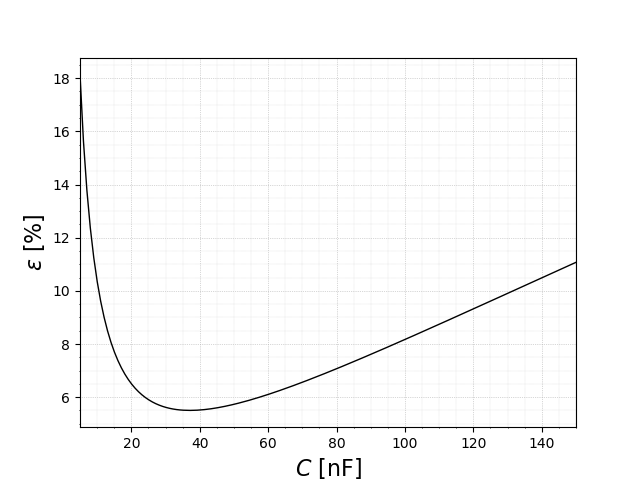}
\caption{The correction $\varepsilon$ as a function of the capacitance $C$.
The behaviour of the correction confirms the analysis made in Sec.~\ref{Sec:Freq}.}
\label{fig:eps}
\end{figure}

Applying Nyquist theorem to a paralelly connected capacitor and resistor with impedance
\be
Z(\omega)=\left( \frac1{R} + \jm \omega C \right)^{-1}, 
\quad \jm=-\im.
\label{Z}
\ee
for the spectral density of the voltage noise we obtain
\be
(U^2)_f = 4 \kb T \frac{R}{1+(\omega R C)^2}.
\ee
This spectral density is amplified by the 3 steps of the amplifier \Eqref{Ups}
\be
(U_\mathrm{amp}^2)_f = |\Upsilon (\omega) |^2 (U^2)_f
\ee
and for the mean square of the amplified voltage we have
\be
\left < U_\mathrm{amp}^2 (t) \right > =
\int_0^\infty |\Upsilon (\omega) |^2  \frac{4 \kb T R}{1+(\omega R C)^2} \frac{\ud \omega}{2 \pi}.
\label{eq:UR}
\ee
In the approximation of frequency independent amplification $\Upsilon(\omega) \approx Y = y_1 y_2 y_3$ we have
\be
\left < U_\mathrm{amp}^2 (t) \right > \approx
\int_0^\infty \!\! Y^2  \frac{4 \kb T R}{1\!+\!(\omega R C)^2} \frac{\ud \omega}{2 \pi} =
Y^2  \frac{4 \kb T R}{4RC}.
\ee
That is why for simplicity we will represent the exact results as the approximation formula
\be
\left < U_\mathrm{amp}^2 (t) \right > = Y^2 \frac{\kb T}{C} \mathcal{Z},
\ee
corrected by the close to unity coefficient
\be
\mathcal{Z} \equiv 1 - \varepsilon \equiv 
\dfrac{\displaystyle \int\limits_0^\infty |\Upsilon (\omega) |^2  \dfrac{\ud \omega}{1+(\omega R C)^2}}{Y^2 \displaystyle \int\limits_0^\infty \dfrac{\ud \omega}{1+(\omega R C)^2}}.
\ee
The small correction $\varepsilon(C)$ as a function of the input capacitances $C$ is graphically presented in Fig.~\ref{fig:eps}.

High school students are freely able to use this figure, while the university students can calculate it numerically.
We do not recommend the analytical calculation of the integral \Eqref{eq:UR}.

After calculating $\left < U_\mathrm{amp}^2 (t) \right >$ we can substitute it in \Eqref{square} and obtain
\be
\left < U_\square \right > = \frac{\left < U_\mathrm{amp}^2 \right >}{U_\mathrm{m}} \frac{R_1+R_2}{R_1}.
\ee
Using the described above correction factor, \Eqref{eq:V2} now reads
\be
V_2 = \frac{ \mathcal{Z} Y^2}{U_\mathrm{m}}  \frac{R_1+R_2}{R_1}
\frac{R_\mathrm{V}}{R_\mathrm{V}+R_\mathrm{av}} \left<U^2(t)\right>,
\ee
with 
\be
\left < U^2 (t) \right > =
\int_0^\infty  \frac{4 \kb T R}{1\!+\!(\omega R C)^2} \frac{\ud \omega}{2 \pi} = \frac{\kb T}{C}.
\ee
Keeping the expression \Eqref{U*} for $U^*$ unchanged, we finally arrive at the correct formula \Eqref{kb} used for the determination of the Boltzmann constant.

\section{Johnson-Nyquist thermal noise and Callen-Welton Fluctuation-Dissipation Theorem}
\label{Apendix_theory}

In order to understand the work of the described set-up, only the equipartition theorem is necessary
and the equipartition theorem is given in the all introductory courses in physics in not only in the universities but also in many high-school textbooks.
Unfortunately speaking about the thermal fluctuation of the voltage people erroneously suppose 
knowledge of the frequency dependent spectral density described by Nyquist and Callen-Welton fluctuation-dissipation theorems. 
These theorems already belong to the extended courses on statistical physics. 
Definitely those theorems cannot be reproduced by at least of 51\% of physics teachers.
That is why in this appendix we will give a new derivation of all those theorems in the level 
corresponding to the introductory courses on physics read for the future physics teachers.
This appendix is oriented for readers willing to understand the work of the set-up, starting from the ideas and notions given in the university courses on statistical physics and nothing is beyond this frame. 

\subsection{Nyquist theorem and its generalizations}

If the Nyquist theorem\cite{Nyquist:28} is derived from the statistical physics methods, it is viewed as an interesting application of the Callen-Welton Fluctuation and Dissipation Theorem (FDT).\cite{Callen:51,LL5}
In the Landau-Lifshtitz course in theoretical physics, both in the 5-th volume Statistical Physics\cite{LL5} and the first edition of the 8-th volume Electrodynamics of Continuous Media,\cite{LL8} a derivation following the original one\cite{Callen:51} is given.
In this approach a Gibbs averaging of the Dirac time-dependent perturbation theory is performed.
The practice in teaching physics however, shows that this derivation meant for professionals is unreproducible by the students paying tuition fees and waiting to receive an educational service;
the degradation of the physics education is global and the authors of the present article are unable to point-out any contemporary textbooks, where FDT is derived.
That is why in this Appendix we give a new re-derivation of the FDT approbated by many recruits of students.

In our derivation of the Nyquist theorem we treat the average energy of an inductance
$\left < E_L \right >$ participating in a high-quality
\be
\mathcal{Q}\equiv\frac{\sqrt{L/C}}{R}\gg 1
\ee
resonance circuit depicted in Fig.~\ref{fig:osc}.
We will compare the result derived by analysis of the influence of the spectral density
of the random noise created by the resistor $(\mathcal{E}^2)_f$ with the result of the Gibbs averaging of the energy
\be
\bar{\varepsilon}=\frac12 \hbar \omega_0 \coth\left (\frac{\hbar \omega_0}{2 \kb T}\right),
\quad \left < E_C \right >= \left < E_L \right >= \frac12 \bar{\varepsilon}
\label{eq:Nyq}
\ee
of an oscillator with resonance frequency $\omega_0=1/\sqrt{LC}$.
The impedance of the circuit 
\be
Z(\omega)=\jm \omega L + R + \frac{1}{\jm \omega C}.
\ee
gives for the modulus of the conductivity $\sigma = 1/Z$
\be
\left| \sigma(\omega) \right|^2=\frac{1}{R^2+\left(\omega L-\frac{1}{\omega C}\right)^2}. 
\ee
The spectral density of the current is given by the spectral density of the voltage
\be
(I^2)_f = | \sigma (\omega) |^2 (U^2)_f,
\ee
which for high-$\mathcal{Q}$ resonance circuit gives
\be
(I^2)_f = \frac{\pi}{2} \frac{(\mathcal{E}^2)_f}{RL} \delta(\omega-\omega_0), \quad
(\mathcal{E}^2)_f=4 R \bar{\varepsilon}.
\label{eq:delta}
\ee
We will prove the Nyquist theorem 
\be
(\mathcal{E}^2)_f=4 R \bar{\varepsilon}.
\label{NyT}
\ee

Expressing the dispersion of the current from its spectral density
\be
\left < I^2 \right > = \int_0^\infty (I^2)_f \frac{\ud \omega}{2 \pi},
\ee
a simple integration of the $\delta$-function in \Eqref{eq:delta} gives
\be
\left < E_L \right > = \frac12 L\left < I^2 \right > = \frac12 \bar{\varepsilon}.
\ee
An analogous consideration of the energy of the capacitor
\be
 \left < E_C \right > = \frac{\left < Q^2 \right >}{2C}, \quad
\left < Q^2 \right > = \! \int_0^\infty \!\! (Q^2)_f \frac{\ud \omega}{2 \pi}, \quad
(Q^2)_f = \frac{(I^2)_f}{\omega^2} \nn
\ee
gives
\be
\left < E_C \right > = \frac12 \bar{\varepsilon}
\ee
in agreement with the virial theorem.
In such a way we have verified that the mean energy of an oscillator
\be
\left < E_C \right >+\left < E_L \right > =
\left <  \frac{Q^2}{2C} +  \frac{LI^2}{2} \right > = \bar{\varepsilon}
\ee
coincides with the Gibbs mean value and we have re-derived the Nyquist theorem \Eqref{NyT} as a consequence of detailed balance principle applied to an electric oscillator.
In the classical case of small frequencies $\hbar \omega_0 \ll \kb T$, which covers almost all electronic applications, we obtain the equipartition theorem
\be
\left < E_C \right > = \left < E_L \right > = \frac12 \kb T
\ee
and low frequency asymptotic of the Nyquist theorem for the spectral density of the noise of a resistor
\be
(\mathcal{E}^2)_f=4 \kb T R.
\ee
One can say that the Nyquist theorem is a consequence of the equipartition theorem and detailed balance principle.

\subsection{Nyquist theorem and its generalizations}

\begin{figure}[h]
\includegraphics[scale=0.4]{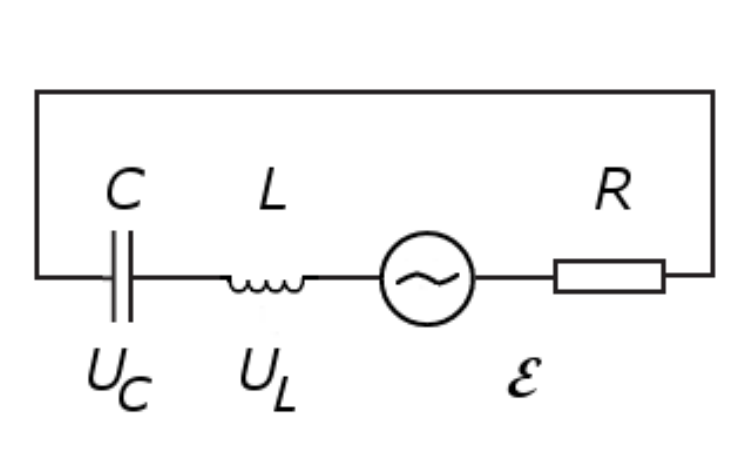}
\caption{Resonance circuit with inductance $Z_L=\jm \omega L$, capacitor $Z_C=1/\jm \omega C$ and resistor $Z_R=R$ creating random electric voltage with spectral density given by Nyquist theorem
$(\mathcal{E}^2)_f=2 \hbar \omega \kb T \coth(\hbar \omega/2 \kb T)$.}
\label{fig:osc}
\end{figure}

In Fig.~\ref{fig:osc} a resonance LC circuit with a resistor $R$ creating random noise is depicted.
The random thermal voltage $\mathcal{E}_\omega$ creates a current amplitude
$I_\omega=\sigma(\omega) \mathcal{E}(\omega)$, where the conductivity $\sigma(\omega)=1/Z(\omega)$ is determined by the total impedance of the sequentially connected inductance, resistor and capacitor
\be
Z(\omega)=\jm \omega L + R + \frac{1}{\jm \omega C}.
\ee
As a gedanken experiment, let us analyse a high quality resonance circuit, for which
\be
R \ll \sqrt{\omega L\frac{1}{\omega C}}=\sqrt{\frac{L}{C}},
\quad \mathcal{Q}\equiv\frac{\sqrt{L/C}}{R}\gg 1.
\ee
In this case the square of the modulus of the conductivity
\be
\left| \sigma(\omega) \right|^2=\frac{1}{R^2+\left(\omega L-\frac{1}{\omega C}\right)^2} \approx
\frac{\pi}{2} \frac{1}{RL} \delta(\omega-\omega_0)
\ee
has a sharp maximum at the resonance frequency $\omega_0 = 1/\sqrt{LC}$ and is negligible far from the resonance and we can use the $\delta$-function approximation 
\be
F(\omega_0)=\int_0^\infty F(\omega)\delta(\omega-\omega_0)\mathrm{d}\omega,
\quad \omega_0>0.
\ee
The coefficient in front of the delta function is given by the integral
\be
\int_0^\infty \frac{\omega^2 \ud \omega}{R^2 \omega^2 + (L \omega^2 -1/C)^2}=
\frac{\pi}{2} \frac{1}{RL},
\ee
which does not depend on the capacitance $C$ and was solved by Schottky~\cite[Eq.~(II)]{Schottky:22}.
Introducing the dimensionless variable $x \equiv \omega L/R$,
and dimensionless parameter $\mathcal{Q}$ the corresponding mathematical problem is 
\be
\mathcal{I}(\mathcal{Q})=
\int_0^\infty \frac{x^2 \ud x}{x^2+(x^2-\mathcal{Q}^2)^2} = \frac{\pi}{2},
\label{eq:intm}
\ee
which can be solved both analytically (given in Subsec.~\ref{subsec:math}) and numerically for different values of $a$.
For $\mathcal{Q}=0$ we have a table integral 
$\int_{-\infty}^{\infty}\mathrm{d}x/(1+x^2)=\pi.$

Using this $\delta$-function approximation, we obtain for the spectral density of the current
\be
(I^2)_f = \frac{\pi}{2} \frac{(\mathcal{E}^2)_f}{RL} \delta(\omega-\omega_0), \quad
(\mathcal{E}^2)_f=4 R \bar{\varepsilon}.
\ee
Experimentally such frequency dependence of the spectral density can be investigated 
using Fourier transformation of digital oscilloscopes and generators. 
Such an equipment is not typical for high-schools and that is why for physics teachers 
the present appendix in only an additional material completing the theory from
general introductory courses. 
Whence for the thermal averaged energies of the capacitor and inductance
\begin{align}
&\left < E_C \right > = \frac12 C \left < U_c^2 \right >, \quad
\left < U_c^2 \right > = \int_0^\infty \left | Z_C\right |^2 (I^2)_f \frac{\ud \omega}{2 \pi},
\nn \\
&\left < E_L \right > = \frac12 L \left < I^2 \right >, \quad
\left < I \right > = \int_0^\infty (I^2)_f \frac{\ud \omega}{2 \pi}, \nn
\end{align}
a trivial integration of the $\delta$-functions
\be
F(\omega_0)=\int_0^\infty F(\omega)\delta(\omega-\omega_0) \ud \omega
\ee
gives
\be
\left < E_C \right >= \left < E_L \right >= \frac12 \bar{\varepsilon}, \quad
\bar{\varepsilon}=\frac12 \hbar \omega_0 \coth\left (\frac{\hbar \omega_0}{2 \kb T}\right).
\ee
The Nyquist result for the spectral density of the noise can be presented in a more general form.

\subsection{Solution of the Schottky integral}
\label{subsec:math}
Contemporary students use Mathematica or Maple but 100 years ago physicists were able 
to perform elementary calculus of integrals.\cite{Schottky:22}
In this subsection we give the calculation of the integral considered by Schottky when he analyzed
the influence of noise the of an reconance circuit.
Let us consider the integrand from \Eqref{eq:intm}
\be
\frac{x^2}{x^2+(x^2-\mathcal{Q}^2)^2}=\frac{C}{x^2+a^2}+\frac{D}{x^2+b^2},
\label{eq:intf}
\ee
where $\pm a$ and $\pm b$ are the $x$ values for which the denominator on the left is equal to 0 and $C$ and $D$ are coefficient we are going to find.
Expanding the divisors both on the left and right sides into a polynomial, we obtain
\begin{align}
&(x^2+a^2)(x^2+b^2)=x^4+(a^2+b^2)x^2+a^2b^2, \\
&x^2+(x^2-\mathcal{Q}^2)^2=x^4+(1-2\mathcal{Q}^2)x^2 + \mathcal{Q}^4.
\end{align}
Both polynomials have to be identical, meaning that the coefficients of the respective $x$ degree terms should be equal.
Clearly, the coefficients of the $x^4$ terms are both equal to 1, and comparing the quadratic and zero degree terms, we obtain a set of two equations (one per each degree term)
\begin{align}
&a^2+b^2=1-2\mathcal{Q}^2 \quad (\mbox{the terms with } x^2), \nn \\
&a^2b^2= \mathcal{Q}^4 \quad (\mbox{the terms with } x^0) .
\label{eq:trans}
\end{align}
From the first equation it is evident that for real values of $a$ and $b$, $\mathcal{Q}<1/\sqrt{2}$, while for 
$\mathcal{Q}>1/\sqrt{2}$ we have $a^2+b^2<0$ and hence $a$ and $b$ are complex.

\subsubsection{Real values of $a$ and $b$, $\mathcal{Q}<1/\sqrt{2}$}
First we are going for a solution in case of real values for $a$ and $b$.
Let us take a closer look at the first of these two equations.
Now going back to Eq.~(\ref{eq:intf}), we combine both fractions on the right hand side into one and the nominator becomes
\be
Cx^2+Cb^2+Dx^2+Da^2=
(C+D)x^2+(Cb^2+Da^2).
\nn
\ee
Comparing it with the nominator on the left hand side $x^2$, we obtain the following two equations
\begin{align}
&C + D = 1 \quad (\mbox{the terms with } x^2),\\
&Cb^2 + D a^2 = 0 \quad (\mbox{the terms with } x^0).
\end{align}
Solving these equations for $C$ and $D$, we get
\be
C=\frac{a^2}{a^2-b^2}, \quad D=-\frac{b^2}{a^2-b^2}.
\ee
Substituting these 	expressions for $C$ and $D$ into Eq.~(\ref{eq:intm}), the integral 
\begin{align}
\mathcal{I}&=\! \int_0^\infty \! \frac{x^2 \ud x}{x^2+(x^2-\mathcal{Q}^2)^2}
=\!\int_0^\infty \! \left ( \frac{C}{x^2+a^2} + \frac{D}{x^2+b^2} \right ) \! \ud x \nn \\
&=\frac{a^2}{a^2-b^2} \int_0^\infty \frac{\ud x}{x^2+a^2}- \frac{b^2}{a^2-b^2} \int_0^\infty \frac{\ud x}{x^2+b^2}. \nn 
\end{align}
The solution of such a table integral is
\be
\int_0^\infty \frac{\ud x}{x^2+a^2}=
\frac{1}{a} \int_0^\infty \frac{\ud (x/a)}{(x/a)^2+1} =
\frac{1}{a}\frac{\pi}{2}
\nn
\ee
and hence for the solution of the integral we obtain
\be
\mathcal{I}=\frac{1}{a^2-b^2}(a-b) \frac{\pi}{2} = \frac{1}{a+b} \frac{\pi}{2}. \nn
\ee
Using the well known binomial theorem $(a+b)^2=a^2+2ab+b^2$, we substitute the values for $a^2+b^2$ and $ab$ from Eqs.~(\ref{eq:trans}) to obtain
\be
(a+b)^2=1-2\mathcal{Q}^2+2\mathcal{Q}^2=1,
\ee
and hence for the solution we have
\be
\mathcal{I}=\frac{1}{a+b} \frac{\pi}{2}=\frac{\pi}{2},
\ee
where we have taken only the positive value.

\subsubsection{Complex values of $a$ and $b$, $Q>1/\sqrt{2}$}

The solution of the integral in this case is analogous.
The parameters $a$ and $b$ are complex.
However for the interesting for us case $\mathcal{Q}\gg 1$
we can recall the analytical continuation.
If $\mathcal{I}(\mathcal{Q})=\pi/2$ for $0<\mathcal{Q}<1/2$
this result has a unique analytical continuation and can be extended from a finite segment to the whole axis $\mathcal{Q}>0.$
Now we address a new derivation of the FDT theorem.

\subsection{Callen-Welton fluctuation-dissipation theorem}

In the spectral density of the voltage noise 
\begin{align}
&(\mathcal{E}^2)_f = 2 R(\omega) \hbar \omega \coth(\hbar \omega/2 \kb T), \\ 
&R(\omega) = \Re(Z(\omega)), \quad Z(\omega)=\frac{1}{\jm \omega C}, \nn
\end{align}
the frequency dependent resistance $R(\omega)$ is the real part of the complex impedance which can be represented also by a frequency dependent capacitance
\be
\omega R(\omega) = \omega \Re \left (\frac{\im}{\omega C(\omega)}\right )=
\frac{\im}{2}(C^{-1}-C^{-1*})=\frac{C^{\prime\prime}}{\left | C\right |^2}, \nn
\ee
where $C^\prime=\Re(C(\omega))$, $C^{\prime\prime}=\Im(C(\omega))$ and $C=C^\prime+\im C^{\prime\prime}$.
Here we suppose an arbitrary frequency of the impedance represented by a generalized capacitance.
On the other hand, the spectral density of the charge is also given by the capacitance
\be
(Q^2)_f = \left | C(\omega) \right |^2 (\mathcal{E}^2)_f.
\ee
In such a way, the Nyquist theorem for the thermal noise can be rewritten as
\be
(Q^2)_f = 2 \hbar C^{\prime\prime}  \coth(\hbar \omega/2 \kb T)
\ee
and for thermal averaged charge fluctuations we finally arrive at
\be
\left <Q^2 \right >\! = \!\! \int_0^\infty \!\!\! (Q^2)_f \frac{\ud \omega}{2 \pi} \!=\!
\frac{\hbar}{\pi} \!\! \int_0^\infty \!\!\! C^{\prime\prime}(\omega)
\coth\!\left(\!\frac{\hbar \omega}{2 \kb T}\!\right) \ud \omega.
\label{eq:chfl}
\ee
We use the spectral density $(\mathcal{E}^2)_f =2(\mathcal{E}^2)_\omega$ which takes into account the folding of positive and negative frequencies  $\omega$.

Let us recall again the formula for the energy when the considered capacitor $C$ is under the voltage
$\mathcal{E}=Q_0/C_0$ created by a large capacitor (charge reservoir) $C_0 \gg C$
\be
E=\frac{Q^2}{2C}+\frac{(Q_0-Q)^2}{2C_0} = \frac{Q^2}{2C}-\mathcal{E}Q + \mbox{const}.
\ee
One can consider that
\be
V=-\mathcal{E}(t)Q
\label{eq:inter}
\ee
is the interaction energy of the capacitor with the voltage source.
The formulae Eqs.~(\ref{eq:chfl}) and (\ref{eq:inter}) are so general that it deserves to change the notations
\be
\hat\alpha(\omega) = C(\omega), \; \hat x=Q, \; f=\mathcal{E}, \; \hat V=V, \;
\left < \hat x \right >_\omega=\alpha_\omega f_\omega. \nn
\ee
Analogously in time representation
\be
\overline{Q}(t)=\int_{-\infty}^{t}C(t-t^\prime)U(t^\prime)\mathrm{d}t^\prime
\ee
becomes
\be
\overline{x}(t)=\int_{0}^{\infty}\alpha(\tau)f(t-\tau)\mathrm{d}\tau.
\ee
In this case the Nyquist theorem can be rewritten as
\be
\left <x^2 \right >=
\frac{\hbar}{\pi} \int_0^\infty \alpha^{\prime\prime}(\omega)
\coth\!\left(\!\frac{\hbar \omega}{2 \kb T}\!\right) \ud \omega, \quad \hat V = -f(t) \hat x.
\ee
cf.~Eq.~(V.124.10) and Eq.~(V.123.1) of the Landau-Lifshitz course of theoretical physics.~\cite{LL5}
The terminology also has to be changed and the Nyquist theorem is now well-known as
fluctuation-dissipation theorem (FDT) for generalized susceptibility $\alpha(\omega)$ giving the general relations between the fluctuations of some quantum variable $\hat x$ and dissipation (absorptive) part of the generalized susceptibility $\alpha^{\prime\prime}=\Im (\alpha(\omega))$.

Callen and Welton rederived the Nyquist FDT applying Gibbs thermal averaging of the quantum mechanical second order perturbation theory, as it is now described in every professionally written textbook on statistical physics.
Even the history of physics is rewritten and in some textbooks one can read that the Nyquist theorem is an interesting application\cite{LL9} of the FDT.
The Sutherland-Einstein relations between the diffusion coefficient and mobility is one of the first examples of he FDT.\cite{Sutherland:05,Einstein:05,Pais:82}

In short we represented all those theorems of the statistical physics as simple consequence of the principle of the detailed balance applied to the simplest physical system -- the harmonic oscillator.
But each creation is a child of its own time, nowadays after more than a century of technology development, every teacher is able to illustrate the thermodynamic fluctuations of the voltage and the equipartition theorem after only a single day work for the building of the experimental set-up.
Tenths of students used the described experimental set-up during the EPO5 and successfully determined the Boltzmann constant and the elementary formulae describing the electronic circuit operation as well.

What has changed since the time of Habicht brothers till now?
That is the invention/appearance of low-noise operational amplifiers such as ADA4898 and multipliers of the AD633 type, which make possible RMS voltages measurements of million times amplified thermal noise.
Now could be motivated to follow our new derivation of the Nyquist and Callen-Welton theorems which is actually detailed balance principle applied to the oscillator written in electric variables.

\end{document}